\documentclass[aip,rsi,reprint,floatfix]{revtex4-1} 

\usepackage{graphicx,sidecap,hyperref}

\begin{document}



\title{A setup for extreme-ultraviolet ultrafast angle-resolved photoelectron spectroscopy at 50-kHz repetition rate}

\author{Jan Heye Buss}
\thanks{These authors contributed equally to this work.}
\affiliation{Materials Sciences Division, Lawrence Berkeley National Laboratory, Berkeley, California 94720, USA\looseness=-1}
\author{He~Wang}
\thanks{These authors contributed equally to this work.}
\affiliation{Materials Sciences Division, Lawrence Berkeley National Laboratory, Berkeley, California 94720, USA\looseness=-1}
\author{Yiming~Xu}
\thanks{These authors contributed equally to this work.}
\affiliation{Materials Sciences Division, Lawrence Berkeley National Laboratory, Berkeley, California 94720, USA\looseness=-1}
\author{Julian~Maklar}
\affiliation{Materials Sciences Division, Lawrence Berkeley National Laboratory, Berkeley, California 94720, USA\looseness=-1}
\author{Frederic~Joucken}
\affiliation{Materials Sciences Division, Lawrence Berkeley National Laboratory, Berkeley, California 94720, USA\looseness=-1}
\author{Lingkun~Zeng}
\affiliation{Materials Sciences Division, Lawrence Berkeley National Laboratory, Berkeley, California 94720, USA\looseness=-1}
\author{Sebastian~Stoll}
\affiliation{Materials Sciences Division, Lawrence Berkeley National Laboratory, Berkeley, California 94720, USA\looseness=-1}
\author{Chris~Jozwiak}
\affiliation{Advanced Light Source, Lawrence Berkeley National Laboratory, Berkeley, California 94720, USA\looseness=-1}
\author{John~Pepper}
\affiliation{Advanced Light Source, Lawrence Berkeley National Laboratory, Berkeley, California 94720, USA\looseness=-1}
\author{Yi-De~Chuang}
\affiliation{Advanced Light Source, Lawrence Berkeley National Laboratory, Berkeley, California 94720, USA\looseness=-1}
\author{Jonathan~D.~Denlinger}
\affiliation{Advanced Light Source, Lawrence Berkeley National Laboratory, Berkeley, California 94720, USA\looseness=-1}
\author{Zahid~Hussain}
\affiliation{Advanced Light Source, Lawrence Berkeley National Laboratory, Berkeley, California 94720, USA\looseness=-1}
\author{Alessandra~Lanzara}
\affiliation{Materials Sciences Division, Lawrence Berkeley National Laboratory, Berkeley, California 94720, USA\looseness=-1}
\affiliation{Department of Physics, University of California at Berkeley, Berkeley, California 94720, USA\looseness=-1}
\author{Robert~A.~Kaindl}
\email[Email: RAKaindl@lbl.gov]{}
\affiliation{Materials Sciences Division, Lawrence Berkeley National Laboratory, Berkeley, California 94720, USA\looseness=-1}


\begin{abstract}
Time- and angle-resolved photoelectron spectroscopy (trARPES) is a powerful method to track the ultrafast dynamics of quasiparticles and electronic bands in energy and momentum space. We present a setup for trARPES with 22.3~eV extreme-ultraviolet (XUV) femtosecond pulses at 50-kHz repetition rate, which enables fast data acquisition and access to dynamics across momentum space with high sensitivity. The design and operation of the XUV beamline, pump-probe setup, and UHV endstation are described in detail. By characterizing the effect of space-charge broadening, we determine an ultimate source-limited energy resolution of 60~meV, with typically 80--100~meV obtained at 1--2$\times 10^{10}$ photons/s probe flux on the sample. The instrument capabilities are demonstrated via both equilibrium and time-resolved ARPES studies of transition-metal dichalcogenides. The 50-kHz repetition rate enables sensitive measurements of quasiparticles at low excitation fluences in semiconducting MoSe$_2$, with an instrumental time resolution of 65~fs. Moreover, photo-induced phase transitions can be driven with the available pump fluence, as shown by charge density wave melting in 1T-TiSe$_2$. The high repetition-rate setup thus provides a versatile platform for sensitive XUV trARPES, from quenching of electronic phases down to the perturbative limit.
\end{abstract}


\maketitle

\section{Introduction}

Angle-resolved photoelectron spectroscopy (ARPES) is a key method to determine the electronic structure of solids in energy and momentum space, yielding essential insights into e.g. superconducting, low-dimensional, and topological phases.\cite{Damascelli2003}  The quest to understand and control the dynamics of solids has further motivated the development of time-resolved ARPES (trARPES).\cite{Bovensiepen2012,Smallwood2016,Zhou2018} By resolving the temporal evolution of quasiparticles and electronic bands across momentum space, trARPES provides powerful new ways to disentangle fundamental dynamics and interactions in the ground state and to reveal the nature of transient phases created far from equilibrium.\cite{Perfetti2007, Graf2011, Wang2013, Sobota2013, Zhang2014, Rameau2016, Parham2017, Rohwer2011, Gierz2013, Cabo2015, Wallauer2016, Bertoni2016, Cilento2018}

Intriguing insights have been obtained by trARPES with ultraviolet (UV) probe pulses around 6 eV, whose efficient generation in nonlinear crystals enables measurements with high sensitivity and energy resolution.\cite{Perfetti2007, Graf2011, Wang2013, Sobota2013, Zhang2014, Rameau2016,Parham2017} However, UV-based ARPES cannot access the important high-momentum band structure near the Brillouin zone edge, which extends to $\approx$1 \AA$^{-1}$ momenta in typical solids. Instead, 6 eV photons result in the emission of Fermi-level electrons with $E_{\rm kin}\!\approx 1.5$~eV kinetic energy, which limits access to in-plane momenta below $k_{||} = \sqrt{2 m_{\rm e} E_{\rm kin}} \sin \theta / \hbar = 0.54$~\AA$^{-1}$ even for a large $\theta = 60^\circ$ emission angle. To overcome these constraints and detect dynamics up to the Brillouin zone edge, extreme-UV (XUV) pulses around 10--50 eV are necessary.\footnote{Smaller photon energies are acceptable for large-unit-cell materials or for detection at  extreme angles. Ultra-high energy resolution laser ARPES has utilized 7--11 eV sources that provide increased bulk sensitivity, see Ref. \onlinecite{Zhou2018} and Y.~He, I.~M. Vishik, M. Yi, S. Yang, Z. Liu, J.~J. Lee, S. Chen, S.~N. Rebec, D. Leuenberger, A. Zong, C.~M. Jefferson, R.~G. Moore, P.~S. Kirchmann, A.~J. Merriam, and Z.~X. Shen, Rev. Sci. Instrum. {\bf 87}, 011301 (2016).} The use of XUV probes also entails access to deeper valence bands, more plane-wave final states, and better isolation from low-energy electrons emitted by multi-photon pump interactions.\cite{Aeschlimann1995}

These considerations have motivated the development of XUV trARPES instruments based on high-harmonic generation (HHG) sources. The first HHG-based trARPES was reported in pioneering work by Haight et al., which allowed for picosecond studies of carrier dynamics in semiconductors and metals at 10--540~Hz repetition rate.\cite{Haight1988, Haight1995} Subsequently femtosecond XUV trARPES with significant improvements in signal averaging was demonstrated by increasing the repetition rate to 1--10~kHz and employing 2D hemispherical electron analyzers for parallel angular detection of electronic dynamics in quantum materials,\cite{Mathias2007, Dakovski2010, Rohwer2011, Frietsch2013, Gierz2013, Cacho2014,Plotzing2016} More recently, XUV trARPES was extended to high repetition rates of several 100~kHz and, via intra-cavity HHG, up to ultra-high repetition rates of 100~MHz.\cite{Chiang2015, Wallauer2016, Cilento2016, Puppin2015, Mills2017, Corder2018}

Sensitive trARPES studies of low-energy electronic dynamics in quantum materials requires that high photon flux and high energy resolution are obtained simultaneously. This is possible, however, only at high repetition rates where the photoelectrons are spread out over many pulses to avoid space charge induced broadening and energy shifts. Conversely, increased repetition rates limit the attainable excitation fluence due to weaker pump pulses, and may result in significant sample heating and re-excitation before a material relaxes back to equilibrium.\cite{Kaindl2000, Gedik2005} For instance, at 100~MHz even $10 \mu$J/cm$^2$ excitation fluence per pulse entails an unacceptable kW/cm$^2$ average power deposited into the sample. This motivates high repetition-rate XUV trARPES in an intermediate regime of $\approx$~30--500 kHz, analogous to conditions utilized in sensitive UV-based trARPES measurements of quantum materials.\cite{Perfetti2007, Smallwood2012}

High-harmonic generation poses additional challenges for trARPES with optimal flux and energy resolution. Traditionally, HHG is driven by few-kHz lasers whose energetic mJ-scale pulses readily provide the $10^{14}$~W/cm$^2$ peak intensities needed for efficient phase-matched conversion. Increased repetition rates entail weaker $\mu$J--scale pulses, requiring tight focusing that makes optimal conversion difficult. Previously, using the output of 50--100 kHz Ti:sapphire amplifiers directly yielded low efficiency HHG with $\approx\!3\!\times\!10^9$~s$^{-1}$ photon flux.\cite{Lindner2003, Chen2009, Heyl2012} High average-power Yb and optical parametric chirped-pulse amplifiers recently boosted HHG to $10^{13}$~XUV ph/s at 100--600 kHz repetition-rate.\cite{Cabasse2012, Puppin2015a, Haedrich2015} Phase-matched HHG, however, generally creates broad harmonics ($\approx$0.3--1~eV) that require spectral selection and narrowing with a complex monochromator.\cite{Frassetto2011, Dakovski2010, Frietsch2013} Here, UV-driven HHG provides a novel path to directly generate narrowband XUV harmonics with high efficiency, which can be isolated using thin metal filters.\cite{Falcao2010, Wang2012, Eich2014, Popmintchev2015, Wang2015} Using intense mJ-scale pulses at 10-kHz repetition rate allowed for trARPES with 150 meV energy resolution with this concept.\cite{Eich2014} Recently, we demonstrated efficient UV-driven HHG in the tight-focusing regime at 50-kHz repetition rate, enabling bright XUV harmonics with $< 72$~meV bandwidth.\cite{Wang2015}

In this paper, we present the realization of an XUV tr-ARPES setup operating at 50-kHz repetition rate, which enables sensitive measurements of ultrafast quasiparticle and band structure dynamics up to the Brillouin zone edge. The design and operation will be described for this instrument, which combines our UV-driven femtosecond HHG source with a sophisticated XUV vacuum beamline and ARPES endstation along with optical pumping capabilities. The energy resolution including the effect of space charge broadening and energy shifts is characterized in detail, which exposes an ultimate resolution without monochromator of 60~meV and yields a reference for choosing suitable photon flux regimes in photoemission studies. Sensitive equilibrium and time-resolved ARPES measurements up to the Brillouin zone edge are performed using the transition metal dichalcogenides TiSe$_2$ and MoSe$_2$ as model systems. With this, the capability for fast ARPES measurements are demonstrated, including equilibrium band mapping, and a 65-fs instrumental time resolution is obtained from the photo-induced dynamics. The {50-kHz} XUV trARPES instrument provides a versatile platform for capturing electronic dynamics across momentum space, from the perturbative regime up to pump fluences that drive photo-induced phase transitions.

\begin{figure*}[ht!] 
\includegraphics[width=16 cm]{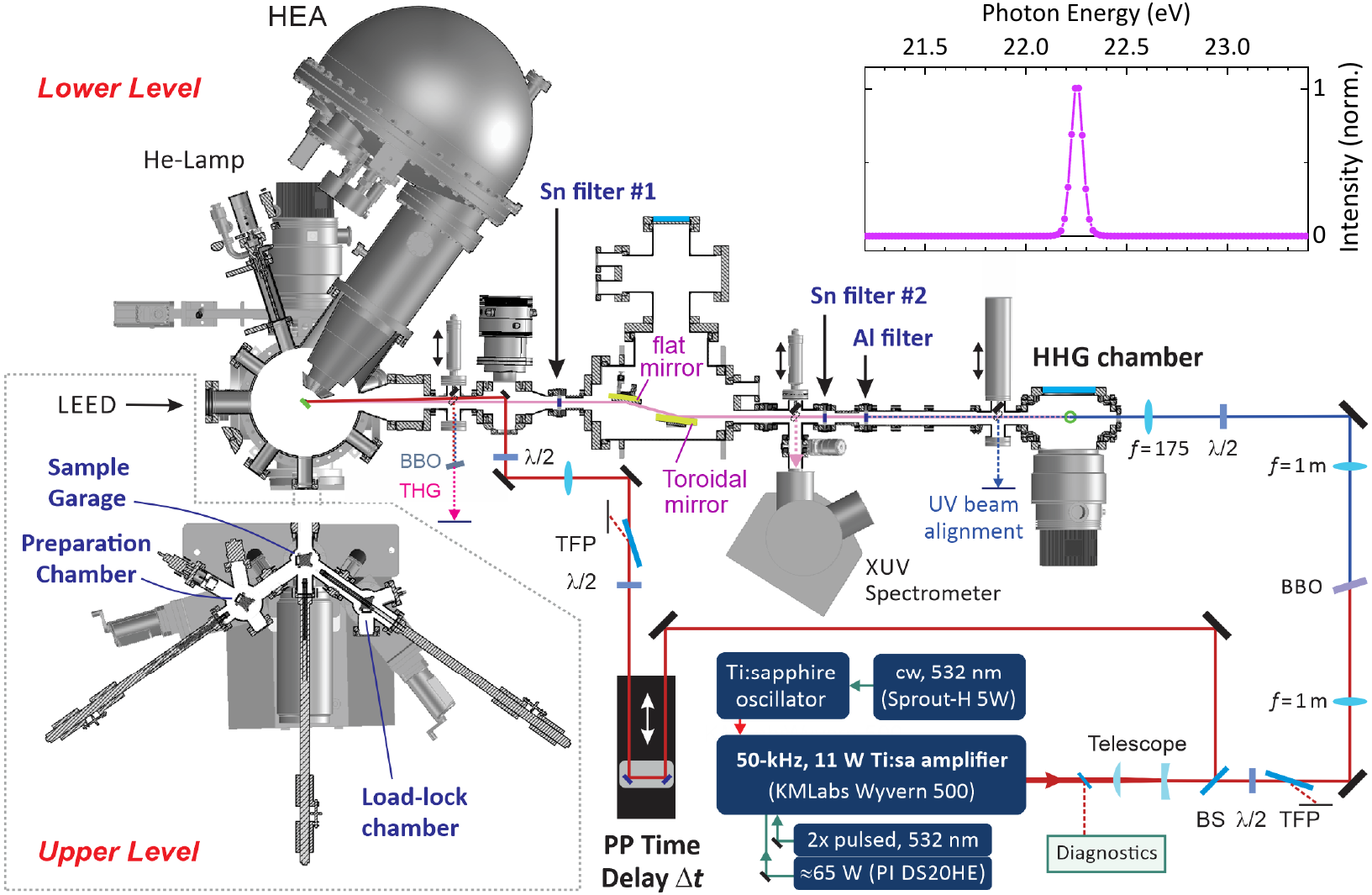}  
\caption{\label{fig1} Technical overview of the extreme-ultraviolet (XUV) time-resolved ARPES setup, including the laser system and optics, XUV beamline with toroidal focusing, and UHV ARPES endstation. Inset: XUV pulse spectrum. Side chambers for sample loading, garage storage, and surface preparation are attached to the upper level of the main chamber, while ultafast and equilibrium ARPES is performed on the lower level with the femtosecond XUV source and a He lamp. TFP: thin film polarizer, $\lambda$/2: half-wave plate, BS: beam splitter, BBO: $\beta$-barium borate, THG: third-harmonic generation, HEA: hemispherical electron analyzer, LEED: low-energy electron diffraction.}
\end{figure*}

\begin{SCfigure*}[][ht!] 
\includegraphics[width=13 cm]{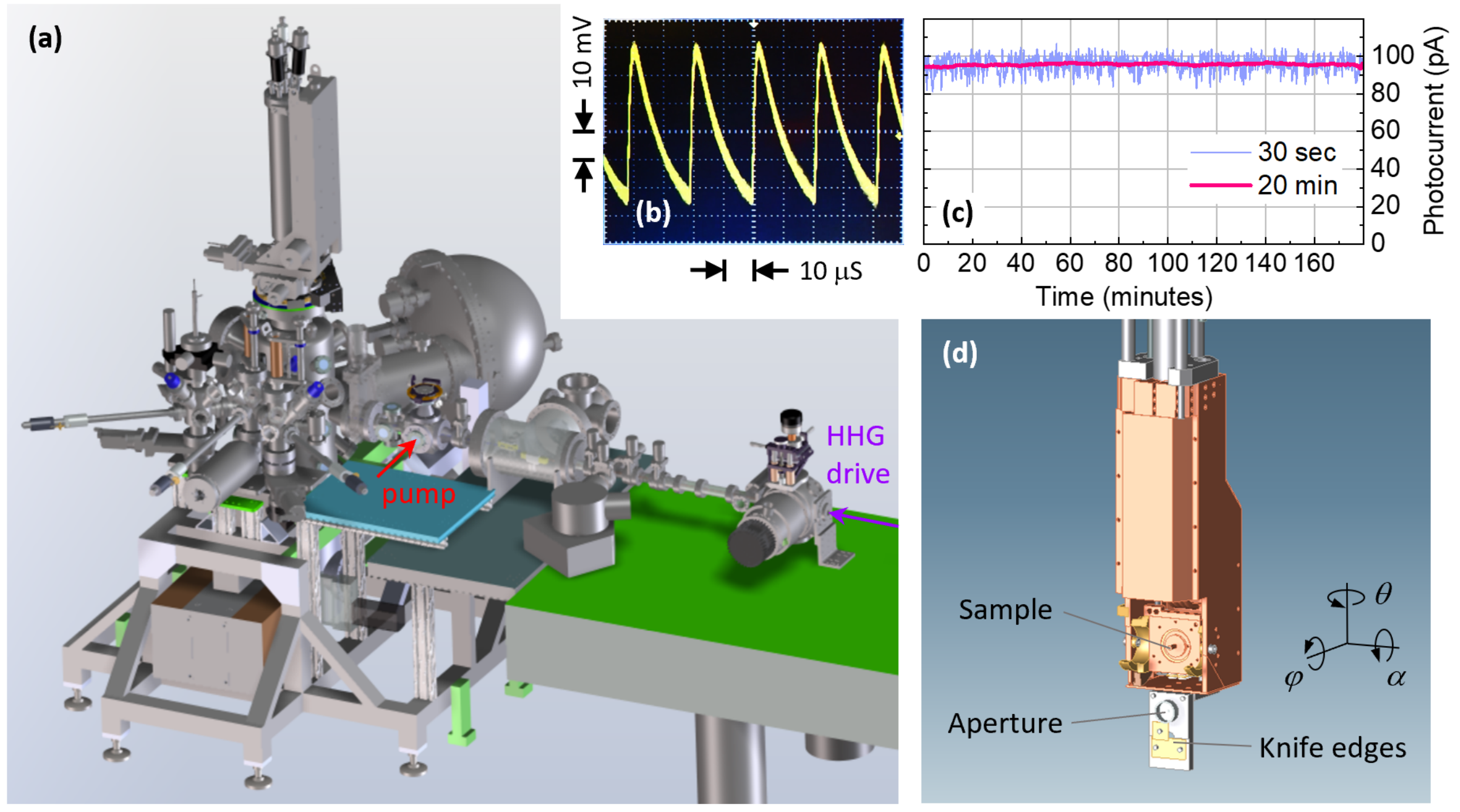}  
\caption{\label{fig2} (a) Side-on view of the ARPES beamline and endstation, with the XYZ manipulator and cryostat mounted from the top. (b) XUV pulse train measured with a Si photodiode. (c) XUV source stability over three hours, tracked via the photocurrent from a Au film and averaged over time windows typical for quick scanning (30~s) and quantitative studies (20~min). (d) Six-axis sample cryostat showing an inserted sample puck, with aperture and knife-edge tools for beam characterization mounted below. The rotational polar ($\theta$), azimuthal ($\varphi$), and tilt ($\alpha$) degrees of freedom are indicated.}
\end{SCfigure*}

\section{EXPERIMENTAL SETUP}
\subsection{Extreme ultraviolet beamline and optical pump capabilities}

Figure 1 shows the layout of the trARPES setup, encompassing the optical configuration for the femtosecond XUV source and pumping capabilities, along with a customized XUV beamline and ultra-high vacuum (UHV) ARPES endstation. At its outset, the ultrafast setup employs a cryo-cooled regenerative amplifier (KMLabs Wyvern 500) to generate near-infrared (near-IR) femtosecond pulses at 50 kHz repetition rate with up to 11~W average power. This amplification stage is seeded by the output of a home-built, 76~MHz Ti:sapphire oscillator, which is pumped by 4.5~W from a green solid-state laser (Lighthouse Photonics Sprout) and is optimized for high pointing stability to minimize spectral drifts in the subsequent pulse stretcher. The amplifier stage, in turn, is pumped by two green, nanosecond pulsed Nd:YVO$_4$ lasers (Photonics Industries DS20HE) with $\approx\!65$~W combined power. After pulse compression, the 50-kHz near-IR amplifier output consists of 50-fs pulses centered around 780 nm wavelength with typically $\approx$200~$ \mu$J pulse energy.

The amplified beam is first passed through a 4:5 lens telescope, reducing it to $\approx\!\!4$~mm width to minimize losses on subsequent optics. The pump beam is separated off with a 30\% beam splitter, with the remaining $130~\mu$J used to generate the XUV harmonics. To enable efficient conversion despite limited pulse energies available at 50-kHz repetition rate, we utilize UV-driven HHG in the tight-focusing geometry. This cascaded scheme boosts the HHG efficiency by two orders-of-magnitude via atomic dipole scaling and improved phase-matching. For this, the near-IR pulses are loosely focused into a 0.5-mm thick $\beta$-barium borate (BBO) crystal to generate 390~nm pulses with $\approx\!48~\mu$J pulse energy. After separating off the fundamental with dielectric mirrors, the UV is then focused tightly into Kr gas -- resulting in the generation of a comb of XUV harmonics with up to $3\times10^{13}$ photons/s in the brightest, 7\textsuperscript{th} harmonic line.\cite{Wang2015}

The XUV linear polarization angle is controlled via the polarization of the UV driving pulses with a half-wave plate before the HHG chamber, yielding sensitivity of the ARPES probe to electronic bands of different orbital character. Moreover, to achieve reproducible HHG conditions small day-to-day laser power variations are manually corrected upstream by a half-wave plate and thin-film polarizer. The XUV beam is generated within a 1-mm inner diameter, end-sealed glass capillary mounted vertically inside a vacuum chamber. The gas pressure is controlled by an automated regulator to maintain optimal phase-matching conditions at $\approx\!60$~Torr backing pressure. Small laser-drilled holes in the capillary sidewall provide beam access, while gas escaping into the chamber is extracted with a 550 L/s turbopump to keep the chamber pressure $<$1.6~mTorr and minimize XUV reabsorption. A high throughput pump (Pfeiffer ATH 500M) is used here to enable extended measurement times $>$16~h without excessive blade heating by the heavy Kr atoms. It should be noted that, as a trade-off for higher flux, the intrinsically lower harmonic cutoff in UV-driven HHG limits the choice of harmonics to optimize for the ARPES cross section or measure different out-of-plane momenta $k_z$ in bulk solids.

As detailed in Fig.~1, an evacuated beamline is connected to the HHG source and allows for the characterization, filtering, and focusing of the XUV beam as it propagates towards the ARPES chamber. The direction of the harmonics is controlled by the incident driving beam, aided by imaging the UV beam after the capillary on a screen, and is highly reproducible in day-to-day operations. The XUV beam diverges slowly with $<6$~mrad full width at half-maximum (FWHM) and is refocused onto the sample using a gold coated toroidal mirror operated under $7.5^\circ$ grazing incidence with 500-mm effective focal length.\footnote{Toroid base radius 7661.298 mm, cylinder radius 130.526 mm (ARW Optical Corp).} An additional grazing-incidence reflection from a motorized, Au-coated flat mirror allows for fine-tuned beam steering onto the sample. This resulted initially in a focal spot size of $162~\mu$m FWHM with, however, unacceptably large space-charge broadening of the ARPES spectra. We therefore increased the distance of the HHG chamber to the toroidal mirror by 10~cm to widen the XUV spot to $\approx\!650~\mu$m FWHM. During optimization, the XUV photon flux is measured with a calibrated Si photodiode (AXUV100G), while the spectrum is monitored on a CCD with an evacuated grating spectrometer (McPherson 234/302).

The inset of Fig. 1 plots the spectrum of the brightest harmonic at 22.25 eV. To select this harmonic, 300-nm thick metal foils are inserted along the beam path. This straightforward isolation is enabled by the large 6.2 eV spacing of the UV-driven harmonics, and avoids both the complexity and temporal broadening inherent to grating monochromators.\cite{Wang2015,Falcao2010,Eich2014} An Al filter located $\approx\!0.7$~m after the source point transmits 7.5\% of the 22.3~eV line, while fully blocking the rapidly diverging residual UV beam and the 3\textsuperscript{rd} and 5\textsuperscript{th} harmonics. In turn, a Sn foil with 8\% transmission at 22.3~eV blocks the weak 9\textsuperscript{th} harmonic at 28.7~eV. The resulting photon flux impinging on the sample is $\approx\!5\!\times\!10^{10}$~ph/s when the transmission through these filters and losses on the optics is taken into account.\footnote{As the 9\textsuperscript{th} harmonic is very weak, $\approx\!6\!\times\!10^{11}$~ph/s XUV flux could be obtained by omitting the Sn filter which, however, is hindered by space charge considerations.} A second Sn filter with twice lower transmission of $\approx\!4$\% can be inserted, allowing for two additional flux combinations of $\approx\!2.5\!\times\!10^{10}$ and $\approx\!2\!\times\!10^9$ photons/s on the sample.

Two 300 L/s turbopumps on the toroidal mirror and subsequent chamber provide for differential pumping with a pressure gradient of 3-4 orders-of-magnitude from the HHG chamber to the main ARPES chamber. The metal filters are mounted on windowed gate valves and provide an additional $\approx$4 orders-of-magnitude isolation each, which is essential to retain the $10^{-11}$~Torr UHV conditions in the ARPES chamber while the light source is running.

The pump beam is sent to the sample after reflection from a gold-coated mirror inside a small vacuum chamber. Its edge is positioned close to the XUV beam path, ensuring a near-collinear propagation of optical pump and XUV probe beams within $\approx\!0.6^\circ$. An $f=750$~mm lens just outside the chamber is used to loosely focus the pump beam onto the sample to typically 1~mm spot diameter. Coarse beam steering occurs by rotating the mirror inside the beam-recombination chamber, while fine tuning of the pump-probe spatial overlap is achieved with a mirror directly before the chamber's entrance window. Before focusing, the pump intensity is variably adjusted with a half-wave plate and thin-film polarizer, while a second half-wave plate controls the polarization. Moreover, the pump pulses can be frequency-doubled in a BBO crystal for UV excitation. The pump-probe time delay $\Delta t$ is scanned up to 1.5~ns with a motorized delay stage. Temporal pump-probe overlap is initially found by sending the near-IR pump and residual UV driving beam (in absence of metal filters) out of the beamline after the beam-recombination chamber and monitoring the sum frequency signal generated in a 200-$\mu$m thick BBO crystal.

\begin{SCfigure*}[][!hb] 
\includegraphics[width=12 cm]{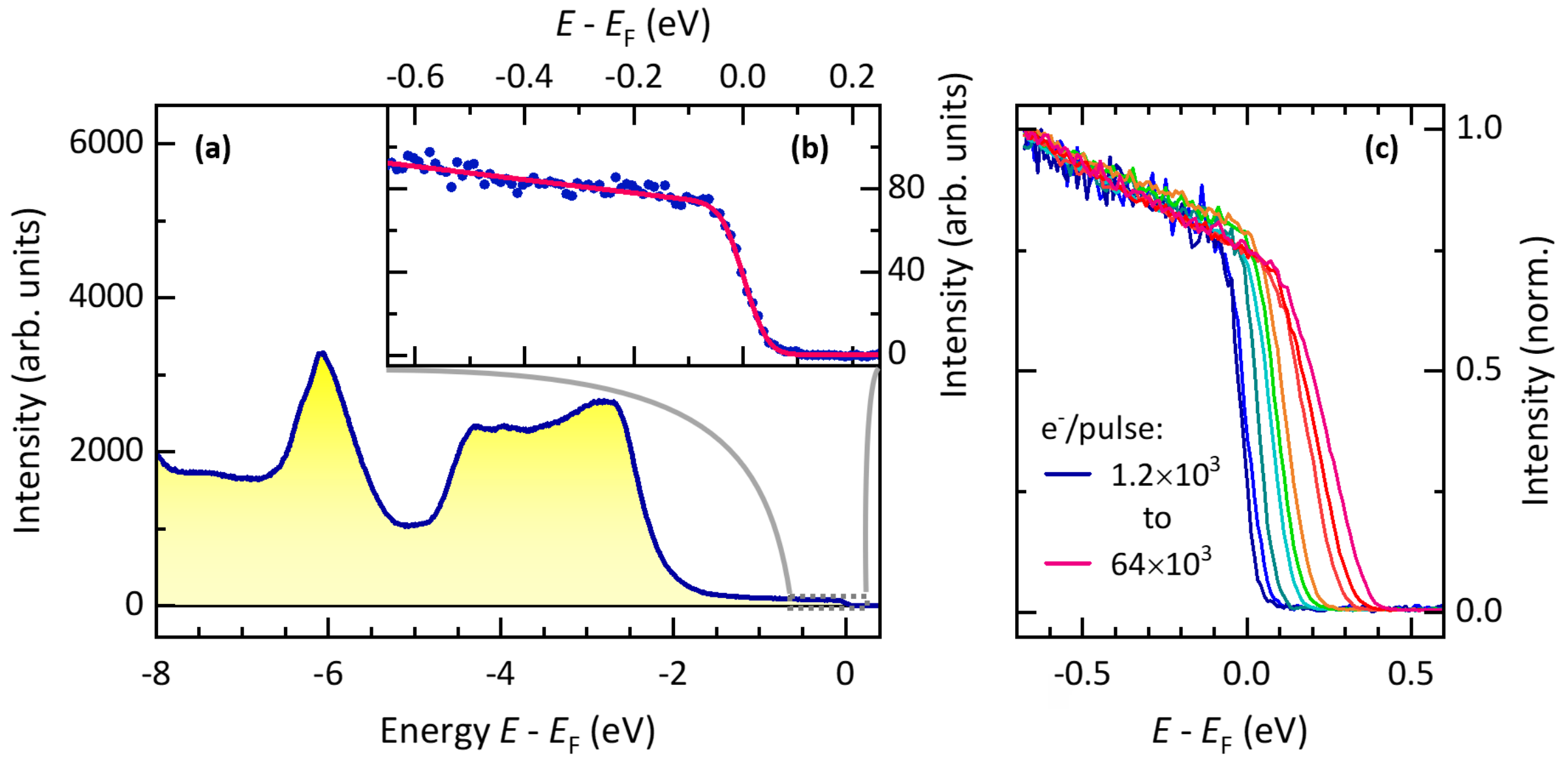}  
\caption{\label{fig3} (a) Energy distribution curve (EDC) measured by  photoemission from Au. (b) Data expanded around the Fermi-edge (dots), and Fermi-function fit convoluted with $\Delta E = 60.4$~meV Gaussian broadening (red line). (c) EDCs at different XUV probe fluences. In all panels, the Fermi energy $E_{\rm F} = 17.68$~eV is subtracted, corresponding to a 4.57~eV work function in good agreement with typical values for gold.\cite{Graf2010}}
\end{SCfigure*}

\subsection{ARPES Endstation}

Time-resolved photoelectron spectroscopy is carried out with a state-of-the art ARPES endstation, the design of which is shown in Figs.~1 and 2(a). The main provides a $5\times10^{-11}$ Torr base pressure for extended trARPES studies, using 500 L/s ion and turbo pumps and a Ti sublimation pump. Photoelectrons are detected with a hemispherical electron analyzer (HEA, Scienta R4000), equipped with a multi-channel plate and CCD detector to monitor the electron distribution in energy and momentum. Rapid data acquisition is facilitated by the analyzer's large 200-mm radius and consequent high electron transmission, and the entrance slit is oriented vertically. For our typical experimental conditions (20~eV pass energy, 0.2~mm slit width) the analyzer provides $\approx\!10$~meV energy resolution and $\approx0.2^\circ$ angular resolution. To minimize external magnetic fields, the ARPES chamber incorporates two nested layers of 3-mm thick $\mu$-metal, yielding $<\!\!1~\mu$T fields at the sample location. The double-layer $\mu$-metal shielding of the R4000 analyzer is closely coupled to the respective main chamber lining, and thin cuffs in a top-hat design extend from the outer lining into the chamber ports.

Three smaller vacuum chambers are connected to the side of the main ARPES chamber, to provide for sample loading, storage and surface preparation (cf. Fig.~1). The load-lock chamber allows for eight samples to be simultaneously evacuated to $<\!\!10^{-9}$~Torr with an 80~L/s turbopump, as well as gold evaporation with a heated filament. Samples mounted on 0.4" diameter copper holders are transported between the chambers with magnetically-actuated linear transfer arms. After initial loading samples are transferred to the sample garage chamber, which is equipped with an ion and Ti sublimation pump and allows for storage of 16 samples in UHV ($8\!\times\!10^{-11}$~Torr). From there, samples are then moved either into the main chamber or a preparation chamber. The latter is pumped to $\approx\!7\!\times\!10^{-11}$~Torr with a 500~L/s turbopump and enables controlled surface preparation using an electron-beam heater and Ar sputter gun.

As indicated in Fig.~1, the main ARPES chamber encompasses two measurement levels. The upper level connects to the garage chamber, allowing for transfer and insertion of a sample puck onto the cryostat stage (Fig. 2d). The sample temperature can be varied from 13-300 K with the liquid He cryostat, mounted on an XYZ-manipulator and rotary seal from the top of the ARPES chamber. Its design provides for six-axis motion, including control of the 3D sample position and its polar angle $\theta$, azimuthal angle $\varphi$, and tilt angle $\alpha$ (Fig.~2d). Moreover, on the upper level samples can be cleaved with a wobble-stick actuated blade, and a compact low-energy electron diffraction (LEED) instrument enables sample surface characterization and rotational orientation. The lower level is dedicated to photoelectron spectroscopy. It includes the electron analyzer, whose lens protrudes into the measurement space, and a port admitting the XUV probe and optical pump beams. The XUV beam and analyzer lens span an angle of $54^\circ$. Moreover, a Helium lamp (Specs UVS 10-35) is attached under the opposite angle to provide energetically-close 21.2~eV photons for equilibrium ARPES characterizations.

Importantly, the sample is grounded through an external connection with a pA-meter to track the photocurrent. This allows for normalization of the ARPES data to long-term drifts of the XUV flux, which is important when comparing datasets measured at different times and for gauging the level of space charge effects. The pulse-to-pulse variations of the XUV pulse train can be monitored on the Si photodiode, as shown in Fig. 2(b). To characterize the long-term stability of the XUV source, we recorded the photocurrent from a gold film. A representative variation over a timespan of three hours is shown as the blue line in Fig. 2(c), with each data point averaged over 30~s. The red line in Fig. 2(c) shows the same data averaged over 20 minutes, corresponding to a typical ARPES accumulation time for one pump-probe delay.

Below the cryostat a metal assembly is mounted (Fig.~2d), which is linked to the sample ground and allows for spatial characterization and overlap of pump and XUV beams. The XUV profile can be determined by scanning the knife edges across the beam while measuring the photocurrent. Similarly, the pump beam is scanned while detecting the transmitted power through a viewport. For efficient alignment of the spatial overlap, a metal aperture (750-$\mu$m diameter) is first centered on the XUV beam which corresponds to the local photocurrent minimum, and the pump beam is then aligned onto the pinhole by maximizing the transmitted power.

\section{ENERGY RESOLUTION AND SPACE-CHARGE EFFECTS}

In the following, we will discuss the energy resolution achievable with the XUV trARPES setup including the influence of detector resolution, high-harmonic width, and space charge. For this, the Fermi edge broadening was studied in photoemission from polycrystalline Au films evaporated in vacuum, with the surface normal facing the analyzer lens. To reduce thermal broadening, the samples were cooled with liquid nitrogen to $T\approx$82~K. Figure 3(a) shows a momentum-integrated photoemission spectrum, measured at low photon flux to minimize space charge effects. Distinct structure from the Au valence band is discerned over several eV, with the Fermi edge cutoff detailed in Fig. 3(b). To obtain the energy resolution $\Delta E$, we fit the Fermi edge with a convolution (red line, Fig. 3b) between a Fermi-Dirac distribution for the given temperature and a Gaussian of FWHM width $\Delta E$. To account for the slow rise below $E_F$, a small linear component is included. For the data in Fig. 3(a-b) the fit results in a broadening $\Delta E = 60.4$~meV.

The resolution $\Delta E$ is determined by the combination of XUV source linewidth $\Delta E$\textsubscript{XUV}, electron analyzer resolution $\Delta E$\textsubscript{DET}, and space charge broadening $\Delta E$\textsubscript{SC}, such that $\Delta E = \sqrt{\Delta E_{\rm XUV}^2 + \Delta E_{\rm DET}^2 + \Delta E_{\rm SC}^2}$. The analyzer resolution is approximated as $\Delta E_{\rm DET} \approx E_{\rm P}\cdot d_{\rm S}/{2R_0}$ for a given pass energy $E_{\rm P}$, mean hemisphere radius $R_0 = 200$~mm, and entrance slit width $d_{\rm S}$. In the above measurement, $E_{\rm P} = 20$~eV and $d_{\rm S} = 0.1$~mm such that $\Delta E_{\rm{DET}} \approx 5$~meV becomes negligible. Moreover, space charge effects were suppressed by performing the measurement at a low photocurrent of 16 pA, corresponding to an electron cloud with $N=2\cdot10^3$ e\textsuperscript{-}/pulse. As shown below, the space charge broadening is negligible ($\Delta E_{\rm{SC}} < 10$~meV) under these conditions. Thus, the broadening must be dominated by the spectral width of the incident XUV pulses. The HHG spectrum (Fig. 1 inset) exhibits a FWHM of 72~meV, from which the spectrometer resolution must be subtracted in quadrature.\footnote{The vacuum CCD spectrometer with 2400~l/mm grating provides $\approx$~0.1~nm spectral resolution, which corresponds to 40~meV energy resolution at 22.3~eV.} This results in an underlying 59.8~meV XUV harmonic width, in excellent agreement with the resolution obtained in the photoemission spectra in Fig.~3(a-b).

Figure~3(c) shows the Fermi edge with increasing sample current, revealing an energy shift and marked broadening. These curves were obtained from different combinations of the Sn and Al filters and fine tuning of the HHG intensity via the UV driving pulse energy. Figure~4 plots the dependence of the Fermi shift and broadening (blue dots) on the total number $N$ of photoelectrons emitted per pulse. The corresponding incident XUV photon flux is also indicated in Fig.~4 (upper axis), based on a quantum yield of 0.06~e\textsuperscript{-}/photon for 22.3 eV photoemission from gold.\cite{Day1981} At the lowest flux, the energy broadening $\Delta E$ in Fig.~4(b) approaches an offset of $\approx 60$~meV, which represents the achievable resolution in the absence of space charge. We can therefore isolate the space charge contribution $\Delta E$\textsubscript{SC} to the overall energy broadening by subtracting in quadrature the XUV linewidth and detector broadening. The resulting $\Delta E$\textsubscript{SC} is plotted as red squares in Fig. 4(b).

\begin{figure}[!ht] 
\includegraphics[width=5.5 cm]{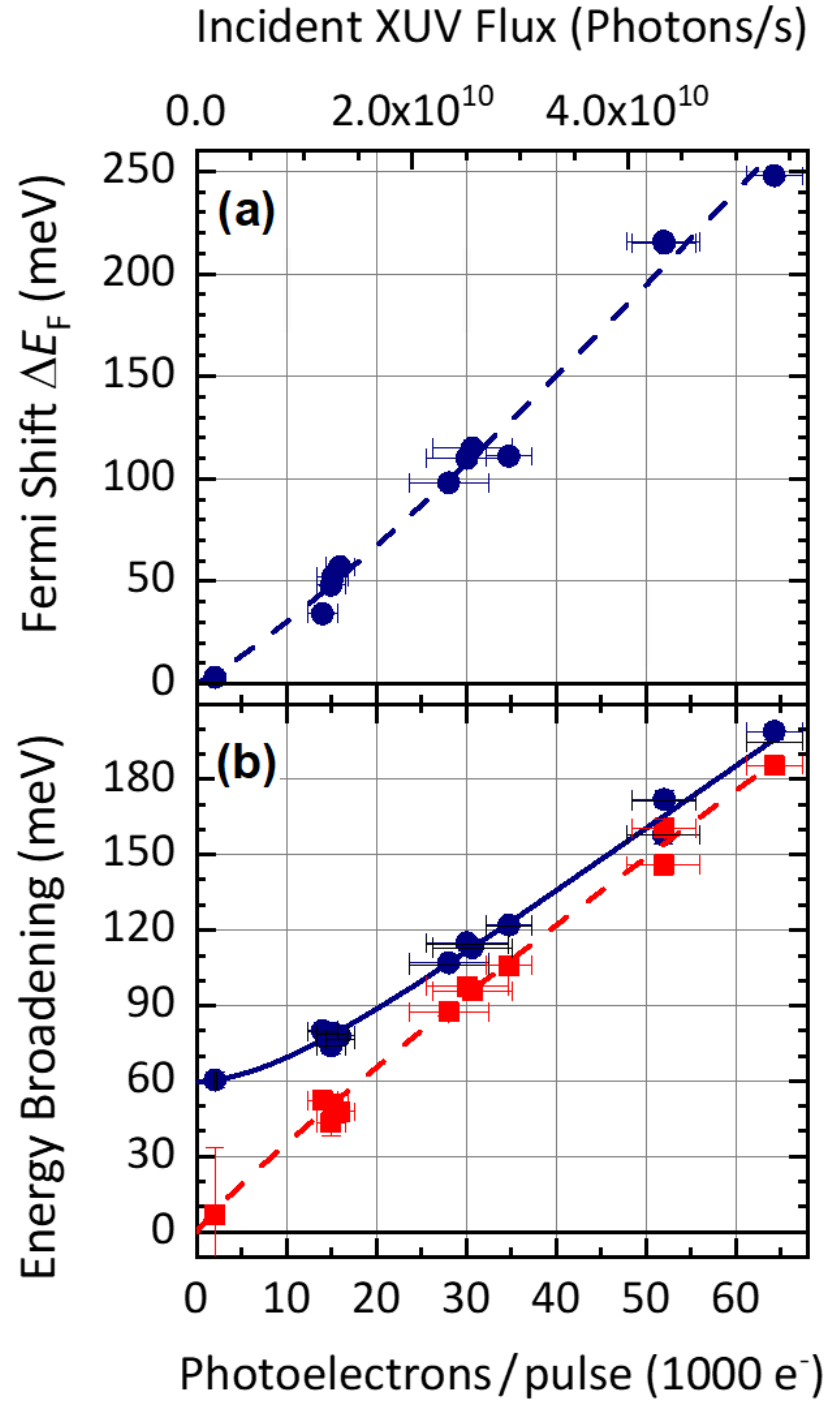}  
\caption{\label{fig4} (a) Space charge induced shift of the Fermi level in Au (dots), along with a power law fit (dashed line) discussed in the text. (b) Total energy resolution $\Delta E$ (blue dots) and space charge contribution $\Delta E_{\rm SC}$ (red squares). Dashed line: power law fit to $\Delta E_{\rm SC}$ as discussed in the text, solid line: total resolution $\Delta E$ with XUV linewidth broadening (59.8~meV) included. Slit widths up to 0.3~mm were used with negligible ($<1.5$~meV) effect on $\Delta E$.}
\end{figure}

The space charge broadening $\Delta E$\textsubscript{SC} and Fermi level shift $\Delta E$\textsubscript{F} in Fig. 4 are of comparable magnitude and exhibit a near-linear dependence on the flux, with a slope comparable to the electrostatic limit.\cite{Hellmann2009} Fits with a power law [dashed lines, Fig. 4(a) and (b)] yield a dependence $\Delta E$\textsubscript{SC}~=~$8.9~\mu{\rm eV}\!\times\!N^{0.9}$ and $\Delta E$\textsubscript{F}~=~$0.7~\mu{\rm eV}\!\times\!N^{1.16}$ on the total number of photoelectrons $N$ per pulse. This behavior observed here in femtosecond photoemission at 22.3 eV confirms the near-linear scaling trend reported in previous experiments and simulations for XUV photoemission,\cite{Zhou2005,Pietzsch2008,Hellmann2009} in contrast to the sub-linear scaling $\Delta E$\textsubscript{SC} $\propto N^{0.5..0.7}$ resulting from low kinetic-energy electrons using 3.1 eV femtosecond two-photon photoemission or a 6 eV picosecond source.\cite{Passlack2006,Graf2010} Mirror charge effects are implicitly included in the experimentally-derived values. Taking into account a velocity of 2.5~nm/fs for 17~eV electrons and nominally $\approx\!30$~fs XUV pulses, the emitted electron cloud has the shape of a flat disc whose thickness is four orders of magnitude smaller than its diameter. The average distance between the electrons is thus independent of the pulse duration and depends only on the XUV spot size,\cite{Hellmann2009}, which varies with the angle of incidence. The exact values will also depend on the photoelectron energy and momentum distribution, i.e. on the material and detector angle. The broadening measured here on Au (Fig.~4) provides a general reference for photoemission with the femtosecond 22.3 eV XUV source, useful for choosing the flux regime for a given experiment.

The measured 60 meV broadening is a remarkably small value for an HHG-based ARPES setup without a monochromator. Due to the 50-kHz repetition rate of the source, experiments with 80 meV resolution can be routinely performed with good photon flux at the sample ($\approx 10^{10}$~ph/s). For experiments where $200$~meV broadening is acceptable, the flux can be further increased to $\approx\!5\times~\!10^{10}$~ph/s to speed up the measurement. Typical experimental data are discussed below.

\section{TIME- AND ANGLE-RESOLVED PHOTOEMISSION SPECTROSCOPY}

In the following, measurements with the 50-kHz XUV photoemission setup are presented which demonstrate its capability to map equilibrium band structures and track ultrafast dynamics with high fidelity up to the Brillouin zone edge. As model systems, we studied the layered transition metal dichalcogenides (TMDs) TiSe$_2$ and MoSe$_2$ which are representative, respectively, of charge-density wave (CDW) and semiconducting compounds. The lamellar structure of TMDs with weak van der Waals interlayer bonding facilitates exfoliation and cleaving, which is not only essential to the fabrication of mono- and few-layer TMD structures \cite{Mak2010} but also yields straightforward access to high quality surfaces for photoemission.

\begin{figure*}[!ht] 
\includegraphics[width=15 cm]{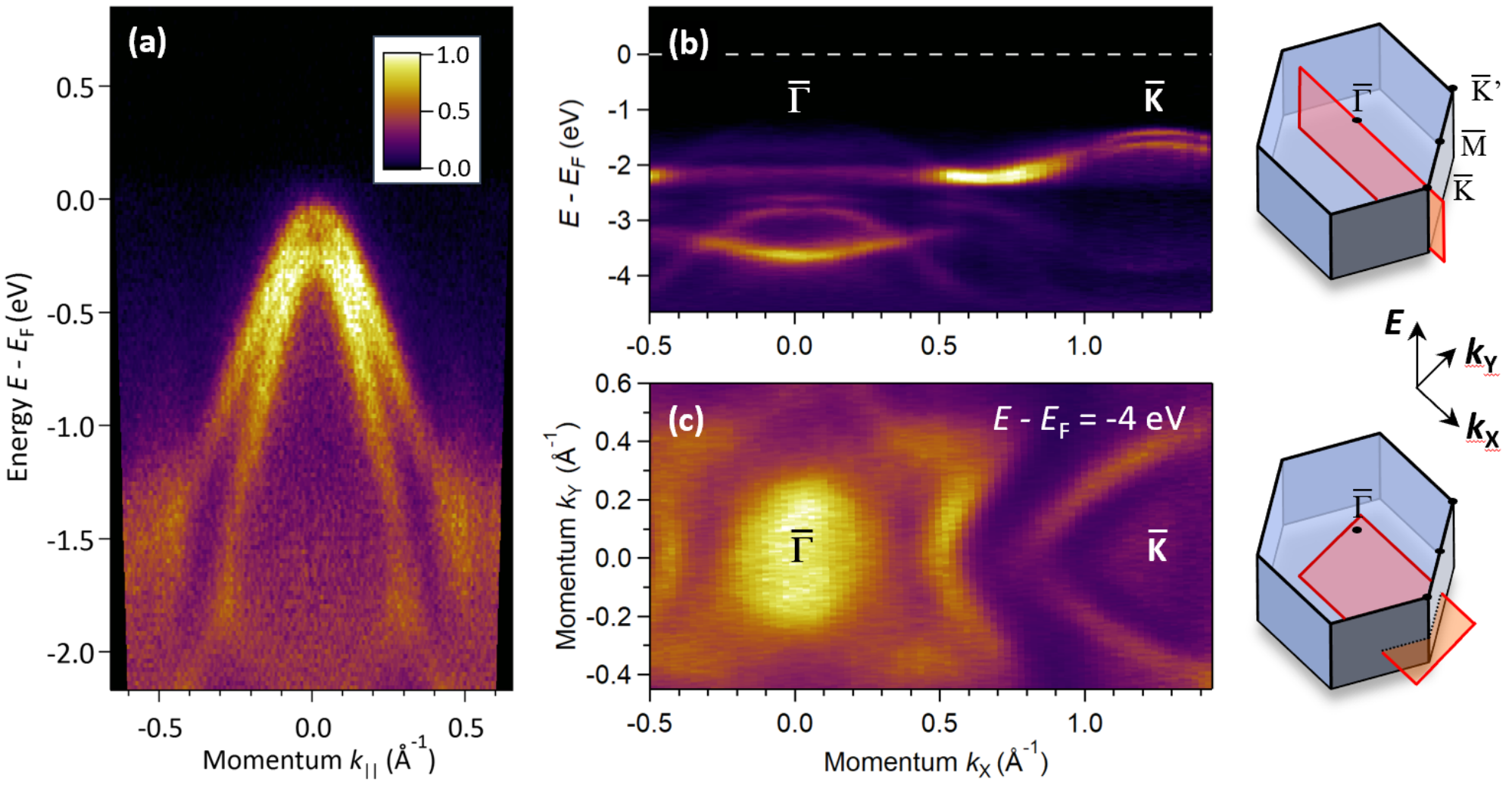}  
\caption{\label{fig5} (a) ARPES intensity map of 1T-TiSe$_2$ measured at $T = 300$~K with the femtosecond XUV source. The data corresponds to 10 sweeps for a total 160~s integration time, obtained at 20~eV pass energy and $\approx$40~pA photocurrent ($N=5\cdot10^3$ e\textsuperscript{-}/pulse). The Fermi energy is $E_{\rm F} = 17.65$~eV, corresponding to a 4.6~eV work function. (b) Band structure of bulk MoSe$_2$ at $T = 86$~K plotted along $\overline{\Gamma}-\overline{\rm K}$, in the notation of the surface-projected Brillouin zone. As illustrated, it represents a cut through a 3D dataset obtained by band mapping. (c) Equipotential cut through the 3D data at fixed energy $E-E_{\rm F} =$~-4~eV.}
\end{figure*}

\subsection{Equilbrium ARPES and Band Mapping}

Figure 5(a) shows an equilibrium ARPES intensity map of 1T-TiSe$_2$ measured using the femtosecond XUV source. The data was taken in the normal phase at room temperature, after cleaving the crystal in-situ. The $30^\circ$ angular range was converted to in-plane momentum $k_{||}$ taking into account the harmonic energy, and states near the zone center are probed with the sample oriented towards the analyzer. Distinct valence bands are directly evident from the ARPES energy-momentum slice, showing two bands dispersing below the Fermi energy which are separated by $\approx\!180$~meV. These features agree well with the spin-orbit split Se~4p bands of TiSe$_2$ observed in high-resolution synchrotron ARPES.\cite{Kidd2002,Rossnagel2011} The ARPES intensity map in Fig.~5(a) was accumulated in only 160~s with a photocurrent of 40~pA ($N=5\cdot10^3$~e\textsuperscript{-}/pulse), which underscores the capability of the XUV trARPES setup to rapidly determine the electronic structure even at the lower end of the flux range of Fig.~4.

Given its sensitivity, the setup can be used not only to measure individual energy-momentum slices but also to map the electronic structure along two dimensions in momentum space and up to the Brillouin zone edge. We performed such measurements on MoSe$_2$, with Fig.~5(b) and (c) showing representative cuts. By combining 71 individual ARPES cuts along $k_y$ taken at different polar angles from $\theta =  -20^\circ$ to $50^\circ$, a three-dimensional (3D) representation of the in-plane band structure was obtained as a function of momenta $k_x$, $k_y$, and energy $E$. With each slice integrated for 180~s, the full map corresponds to 142~minutes total accumulation time. For this, the long-term source stability with only minor intensity drifts over several hours was essential.

Figure~5(b) shows the MoSe$_2$ band structure along the $\overline{\Gamma}-\overline{\rm K}$ direction, representing a cut through the 3D energy-momentum dataset (for the full data, see Supplementary Movie S1). The dispersion of multiple valence bands is clearly evident at energies down to 5~eV below the Fermi level and up to high momenta of $1.5$~\AA$^{-1}$ that exceed the Brillouin zone edge. This band structure measured with the XUV harmonic source agrees closely with previous synchrotron ARPES studies and ab-initio calculations.\cite{Boeker2001,Zhang2014a} Around the $\overline{\rm K}$  point, the known $\approx\!200$~meV spin-orbit splitting is distinctly evident. Moreover, the ARPES intensity at the valence band maximum around $\overline{\Gamma}$ is suppressed by matrix element effects due to the chosen $s$-polarization for the XUV beam. In turn, Fig.~5(c) shows a cut through the 3D dataset at a fixed binding energy of $E-E_{\rm F} = -4$~eV. This equipotential ARPES intensity map clearly evidences the symmetries inherent in the MoSe$_2$ band structure and underscores the ability to easily access the key symmetry points with the 22.3~eV femtosecond probe.

\subsection{Quasiparticle Dynamics}

To gauge the capability for probing non-equilibrium quasiparticles at the Brillouin zone edge, we carried out trARPES measurements of MoSe$_2$ at low pump fluences. For this, near-IR pulses from the amplifier output were used to excite the sample near its A-exciton peak, which drives momentum-direct transitions coupling valence to conduction band states around the $\overline{\rm K}$ point. Figure 6(a) shows the transient ARPES intensity around $\overline{\rm K}$ obtained at low temperatures ($T = 83$~K) for a pump-probe delay $\Delta t = 35$~fs after excitation. A distinct photoemission signal is evident 1.6~eV above the spin-orbit split valence band. Figure~6(b) shows corresponding EDCs, representing the ARPES intensity integrated over momenta within the dashed lines in Fig.~6(a). A peak around $E_{\rm kin} = 18$~eV appears after excitation ($\Delta t = 35$~fs), while it is absent for negative delays ($\Delta t =$~-130~fs) with only a broad background observed. In the measurements the pump spot diameter was $\approx\!900~\mu$m and the sample was driven with $68~\mu$J/cm$^{2}$ incident fluence, which corresponds to $F_{\rm abs}~\approx\!33~\mu$J/cm$^2$ absorbed fluence after taking reflective losses into account.\cite{Beal1979} At this excitation level, the transient peak is easily detectable as a small signal of only 1\% of the valence band intensity.

\begin{figure}[ht!] 
\includegraphics[width=8.5 cm]{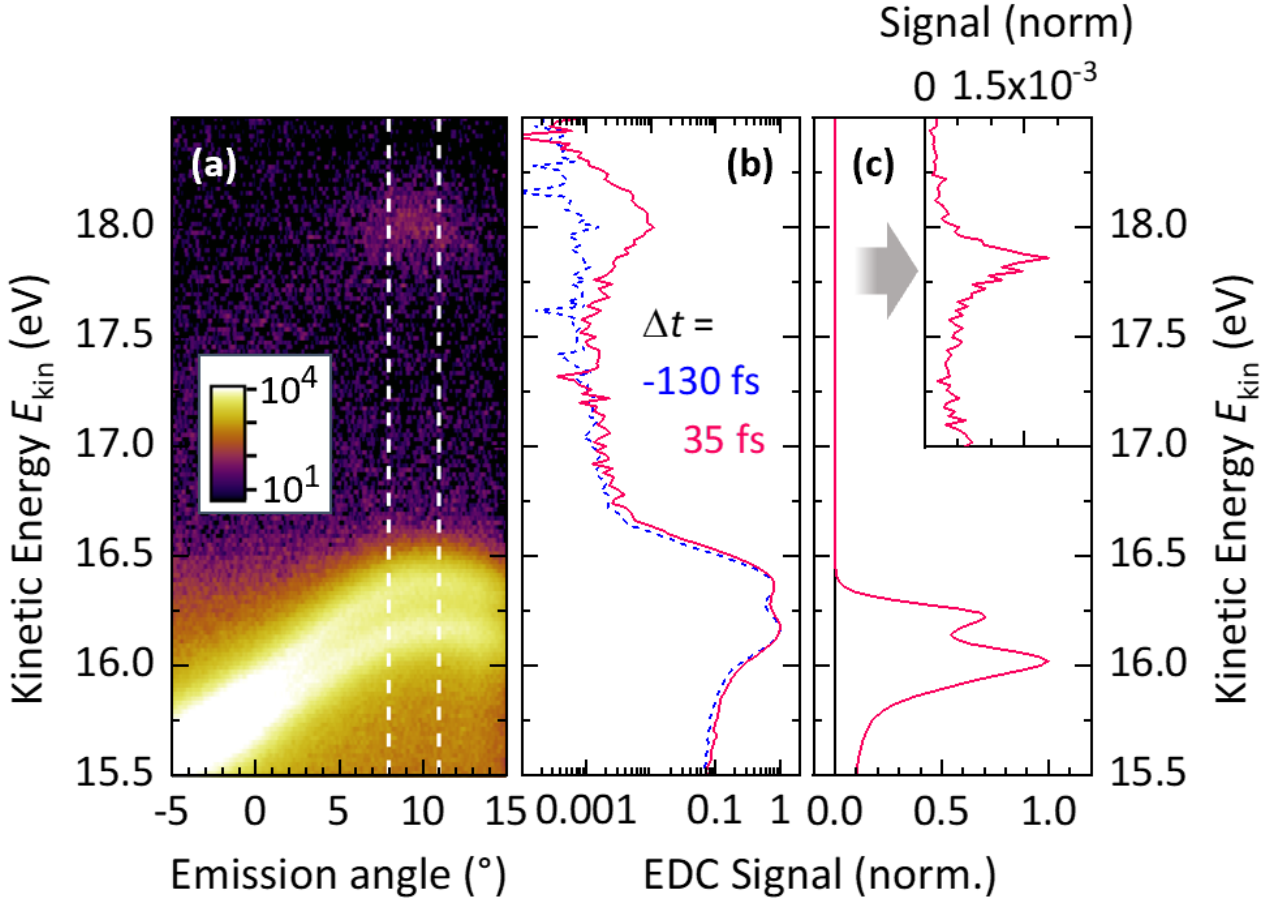}  
\caption{\label{fig6} Time-resolved XUV ARPES of MoSe$_2$ at $T = 83$~K. (a) ARPES intensity map around $\overline{\rm K}$, measured 35~fs after near-IR excitation with absorbed fluence $F_{\rm abs}~=~33~\mu$J/cm$^2$. The color scale is logarithmic. (b) Corresponding EDC on a log-scale, for two different pump-probe delays $\Delta t$ as indicated. The curves correspond to the integration range indicated in panel (a) by the dashed lines. (c) Transient EDC for a very low fluence $F_{\rm abs} = 8~\mu$J/cm$^2$, plotted on a linear scale. Inset: zoom-in evidencing a very small signal due to the perturbatively excited quasiparticles.}
\end{figure}

Importantly, the high repetition rate enables measurements at even lower excitation levels, as shown in Fig.~6(c) for an absorbed fluence of $F_{\rm abs} = 8~\mu$J/cm$^2$. This corresponds to a photoexcited electron-hole pair density of $n_{\rm 2D} = 3\!\times\!10^{11}$~cm$^{-2}$ per monolayer at the surface, after taking into account the pump absorption depth for MoSe$_2$.\cite{Beal1979,Li2014} The trARPES intensity map underlying this EDC was accumulated for 36 minutes around the $\overline{\rm K}$ point with a photocurrent of 80~pA ($N=10^4$ e\textsuperscript{-}/pulse). As evident from the inset of Fig. 6(c), the photo-induced peak is still clearly resolved and corresponds to a signal of only $\approx\!1.7\!\times\!10^{-3}$ of the valence band intensity. Since the noise level is $\approx\!2\!\times\!10^{-4}$ relative to the valence band intensity (Fig.~6c inset), even smaller signals can in principle be detected. Transient pump-induced broadening and energy shifts occur, as seen by comparing the valence band peaks in Figs.~6(b) and (c). They arise from pump-induced space charge or transient surface photovoltage effects,\cite{Oloff2016} which become negligible at the lowest excitation fluence. The data in Fig.~6 thus demonstrate a high sensitivity and underscore the advantage of the 50-kHz setup in accessing perturbative quasiparticle dynamics up to the Brillouin zone edge.

\begin{figure}[hb!] 
\includegraphics[width=7 cm]{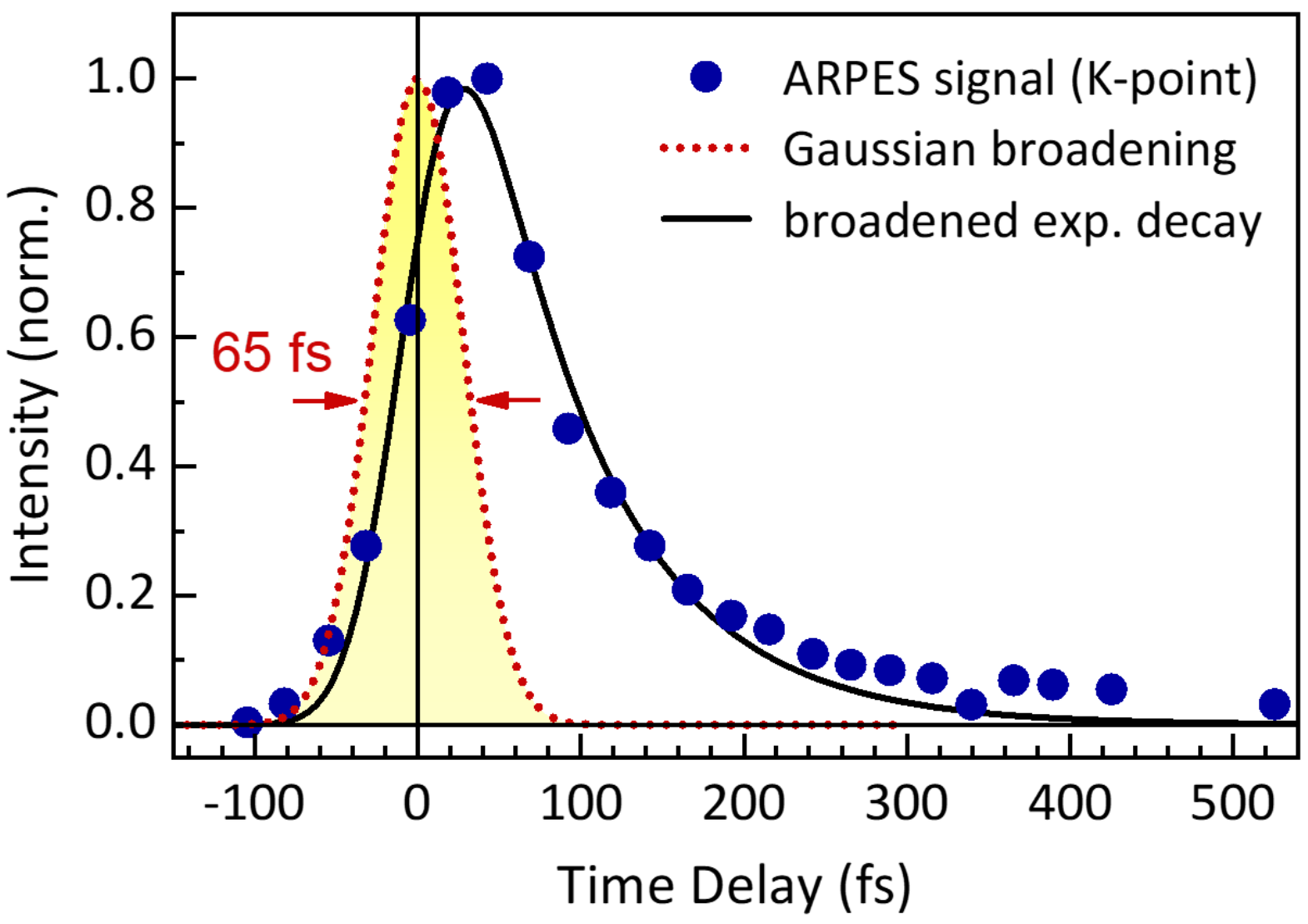}  
\caption{\label{fig7} Dynamics of the photoexcited electron distribution in MoSe$_2$ (dots), obtained by integrating the transient ARPES peak around $\overline{\rm K}$. The data is normalized, with the steady-state background subtracted, and was measured at $T = 85$~K with $F_{\rm abs} = 65~\mu$J/cm$^2$ fluence. Solid line: single-exponential decay $\Phi(t) \cdot \exp{(-t/\tau_{\rm D})}$ convoluted with a broadening of width $\Delta t_{\rm res}$ (FWHM), where $\Phi(t)$ is the Heaviside step function simulating an instantaneous sample response. The fit corresponds to $\Delta t_{\rm res} = 65$~fs time resolution and $\tau_{\rm D} = 77$~fs decay. Dotted: Gaussian time resolution broadening, representing the cross correlation between optical pump and XUV probe pulses.}
\end{figure}

Since resonant pumping promotes the carriers almost immediately into the unoccupied states near the Fermi level, we can evaluate the risetime of the non-equilibrium signal in Fig.~6 to obtain an upper bound on the time resolution of the trARPES setup. The dynamics is plotted in Fig.~7, obtained by integrating the transient peak at $\overline{\rm K}$ in energy and momentum for different pump-probe time delays $\Delta t$. The steady-state background was subtracted, and the signal was measured  at $65~\mu$J/cm$^2$ absorbed fluence for increased signal-to-noise. As evident from Fig. 7, the resulting dynamics exhibits a rapid rise followed by a relaxation on a time scale of several 100~fs, analogous to measurements on other semiconducting TMDs.\cite{Cabo2015,Wallauer2016,Bertoni2016} The solid line is a fit with a single-exponential decay convoluted with a Gaussian broadening function, yielding a time resolution $\Delta t_{\rm res} = 65$~fs. Due to the nearly collinear propagation of the pump and probe beams towards the sample, geometric time smearing is minimized and adds less than 2~fs to this value.\cite{Taft1996} Instead, the 65-fs resolution is fully accounted for by the cross correlation between the XUV probe with an estimated $\approx\!30$~fs transform-limit duration and the near-IR pump whose duration is $\approx\!58$~fs after taking into account the dispersion of 8~m air and $\approx\!9$~mm glass in the setup. \footnote{Accordingly, we expect enhanced time resolution if the pump pulses are shortened using a re-compressed white light continuum or nonlinear optical parametric amplifier. The XUV pulse duration could then also be determined more directly in a cross-correlation measurement, aside from more elaborate gas-based schemes described e.g. by Y. Mairesse and F. Qu\'{e}r\'{e}, Phys. Rev. A {\bf 71}, 011401R (2005).}

\begin{figure*}[!ht] 
\includegraphics[width=15.8 cm]{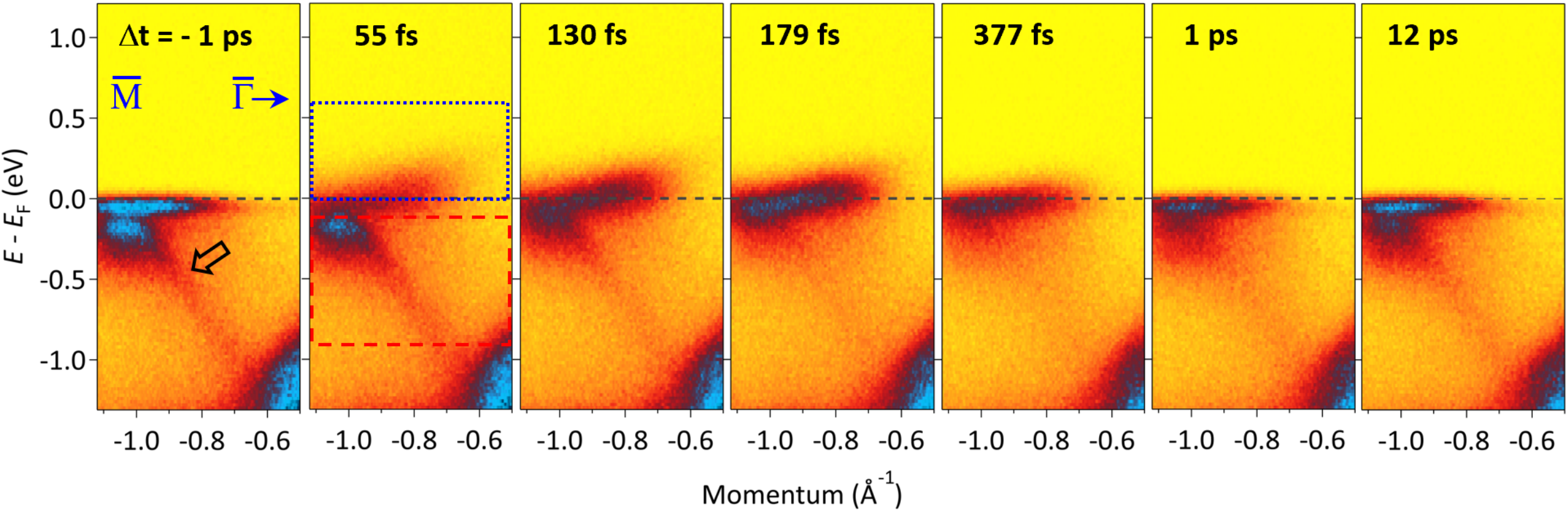}  
\caption{\label{fig8} Time-resolved ARPES intensity maps along the $\overline{\rm M} - \overline{\Gamma}$ direction of 1T-TiSe$_2$, for different time delays $\Delta t$ after excitation with fluence $F_{\rm abs} = 82~\mu$J/cm$^2$. The signals were measured with $\approx$100~pA photocurrent ($N=1.2\cdot10^4$ e\textsuperscript{-}/pulse). Integration areas to evaluate free-carrier and CDW dynamics are indicated, respectively, by the blue dotted and red dashed boxes.}
\end{figure*}

\subsection{Electronic Structure Dynamics}

To demonstrate the capability for detecting photo-induced phase transitions, we measured the dynamics of 1T-TiSe$_2$ as a model CDW compound. It exhibits a phase transition from a metallic to a charge-ordered phase below the transition temperature $T_{\rm CDW} = 202$~K.\cite{Holy1977} The density wave entails strong modifications of the electronic structure, which was previously exploited in trARPES studies of photo-induced CDW melting at lower repetition rates.\cite{Rohwer2011,Monney2016,Mathias2016} To track the temporal evolution with our 50-kHz setup, the TiSe$_2$ crystal was cooled to $T = 83$~K well below the phase transition and photo-excited with the 780-nm wavelength pump pulses.

Figure~8 shows the resulting transient ARPES maps near the high-momentum $\overline{\rm M}$-point for different time delays $\Delta t$ after photo-excitation. Before the arrival of the pump pulse \mbox{($\Delta t =$~-1~ps)}, two bands are evident in the unperturbed slice of the band structure, which is plotted along the $\overline{\rm M} - \overline{\Gamma}$ direction. The signal near the Fermi level corresponds to the conduction band minimum with Ti-3d character, and represents residual background carriers. Below that a dispersing band is evident (thick arrow), which appears only below $T_{CDW}$ and is explained by CDW-induced back folding of the Se-4p valence band at $\overline{\Gamma}$ (cf. Fig. 5a) to the $\overline{\rm M}$-point. This distinct feature thus acts as a spectral marker for the presence of CDW order. Due to the high energy resolution of the XUV probe, the gap between the Ti-3d and back-folded Se-4p bands is clearly resolved at $\overline{\rm M}$.

With increasing time delay, strong changes of the electronic structure are evident in Fig.~8. The pump is set to an absorbed fluence of $F_{\rm abs} = 82~\mu$J/cm$^2$. Upon excitation and shortly thereafter \mbox{($\Delta t =$~55~to~179~fs)}, the ARPES maps reveal a strong quench of the back-folded band at $\overline{\rm M}$, while the Ti-3d conduction band is populated up to higher energies with hot electrons. This directly evidences the melting of the CDW order on an ultrafast time scale, which proceeds further until \mbox{$\approx$350~fs} and then recovers on a ps time scale. Moreover, the gap between the Se-4p and Ti-3d bands is suppressed as the weakened back-folded band merges with the conduction band.

Figure~9 plots the corresponding dynamics of transient carriers and back-folded CDW band intensity. The CDW signal is obtained by integrating the ARPES intensity within the red dashed area indicated in Fig.~8, while the carriers are integrated in the blue dotted area above $E_{\rm F}$. The carriers (dots, Fig. 9b) exhibit a fast rise peaking around $\approx$120 fs, where the delay can be explained by carrier multiplication across the reduced gap.\cite{Mathias2016}. Importantly, the CDW signal (dots, Fig.~9a) is additionally delayed with respect to the carriers and reaches its maximum quench $\approx$200~fs later, underscoring the non-thermal nature of the transient phase. The dynamics can be modeled by an expression $\Delta I(t) \propto \{1-\exp(-t/\tau_{\rm R})\} \cdot \exp(-t/\tau_{\rm D})$, where $\tau_{\rm R}$ and $\tau_{\rm D}$ are the rise and decay time, respectively. A corresponding fit to the CDW signal shown as lines in Fig. 9(a) and (b) yields $\tau_{\rm R}=146$~fs as the quench time of the long-range CDW order.

\begin{figure}[!ht] 
\includegraphics[width=8.4 cm]{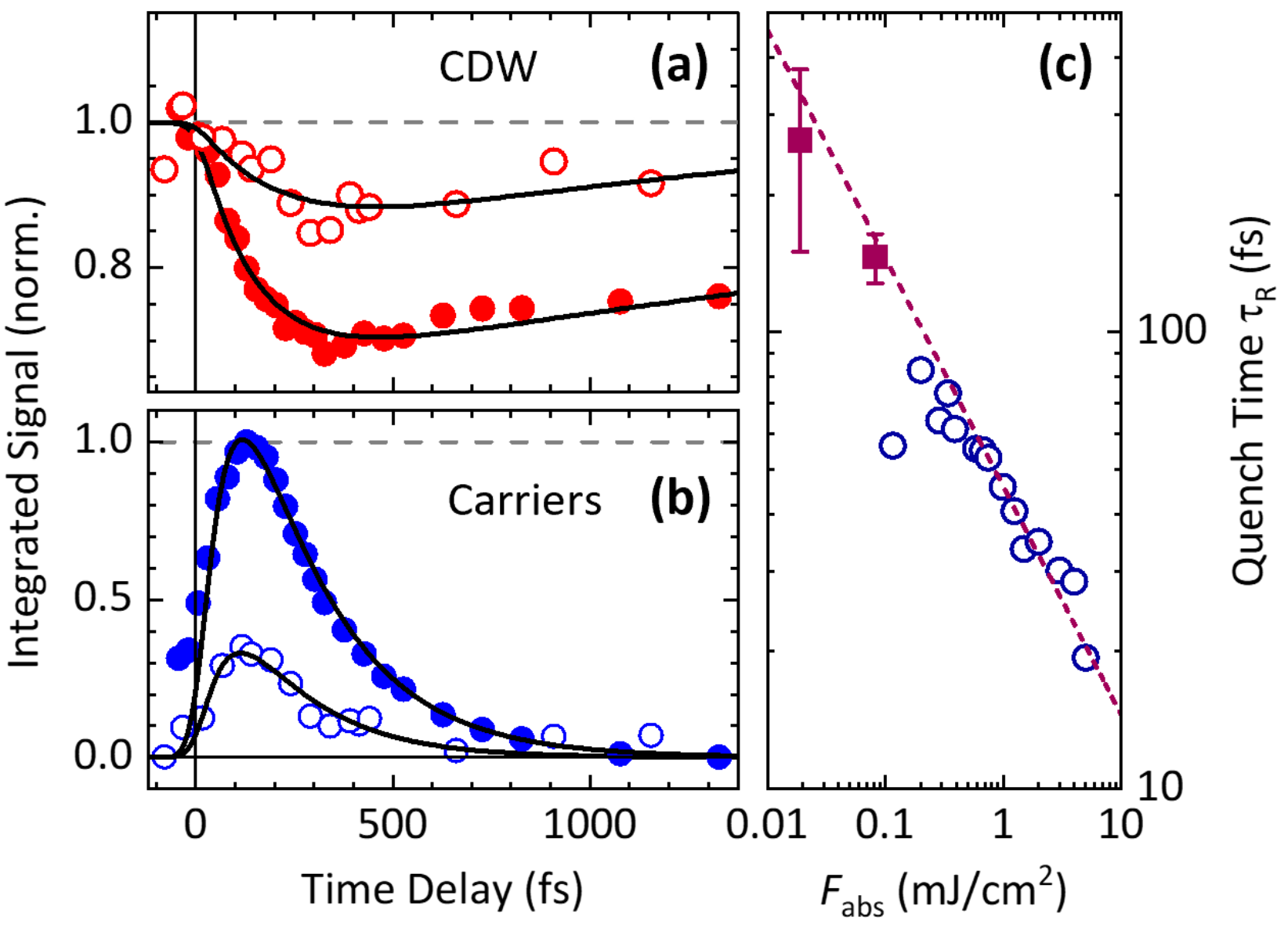}  
\caption{\label{fig9} Carrier dynamics and CDW melting in 1T-TiSe$_2$, represented by the integrated signals of (a) the back-folded CDW band and (b) free carriers in the conduction band near $\overline{\rm M}$. Data are shown for absorbed pump fluences $F_{\rm abs} = 82~\mu$J/cm$^2$ (dots) and $19~\mu$J/cm$^2$ (circles), along with fits as discussed in the text. (c) Quench times from the fits (squares) and from Ref.~\onlinecite{Rohwer2011} (circles), compared to a $\sqrt{F}$ law (dashed).}
\end{figure}

Given the sensitivity of the 50-kHz trARPES setup, we extended the measurements also to a lower fluence of $F_{\rm abs} = 19~\mu$J/cm$^2$. The dynamics [circles, Fig. 9(a) and (b)] reveals a partial melting and reduced carrier signal, with a longer CDW quench time ($\tau_{\rm R}=263$~fs) obtained from the fit. The observation of partial suppression agrees with optical-pump THz-probe studies that showed the full melting of the electronic CDW component only above $\approx\!40~\mu$J/cm$^2$.\cite{Porer2014} The quench times observed here extend the previous trARPES studies towards lower fluences (Fig. 9c), compatible with an approximately square-root scaling with fluence which indicates the key role of screening in melting the electronic order.\cite{Rohwer2011} The results obtained here thus demonstrate the capability of the \mbox{50-kHz} setup to sensitively track dynamics of electronic structure and provide the basis for further explorations of photo-driven phase transitions including the low-fluence crossover regime.

\section{SUMMARY}

In conclusion, we have presented the detailed design and operation of a setup for ultrafast time-resolved ARPES with XUV pulses at 50-kHz repetition rate, enabling sensitive measurements of ultrafast quasiparticle and band structure dynamics up to the Brillouin zone edge. The instrument combines an efficient 22.3 eV femtosecond source with a sophisticated XUV beamline and ARPES endstation along with optical pumping capabilities. The effect of space charge broadening on the energy resolution was characterized up to a flux of $\approx\!5\!\times\!10^{10}$~ph/s on the sample. An ultimate resolution of 60~meV limited by the harmonic width was attained at $\approx\!2\!\times\!10^9$~ph/s, while efficient ARPES with $\approx$80--100~meV resolution is enabled by a higher flux of $\approx$1--2$\times 10^{10}$~ph/s on the sample. To demonstrate the capabilities of the 50-kHz setup, both steady-state and time-resolved ARPES was carried out on the prototypical transition metal dichalcogenides TiSe$_2$ and MoSe$_2$. Fast acquisition was shown to enable full band mapping in equilibrium. Transient ARPES signals were observed in MoSe$_2$ for absorbed pump fluences down to $8~\mu$J/cm$^2$, validating the capability to detect signals as small as $\approx\!10^{-3}$ of the valence band intensity. The risetime of the photo-induced signals revealed a 65~fs instrumental time resolution. In turn, the setup provides sufficient pump fluences to drive photo-induced phase transitions, as confirmed by observing charge density wave melting in 1T-TiSe$_2$. The 50-kHz XUV trARPES instrument thus enables studies of electronic dynamics up to the Brillouin zone edge with with high sensitivity and energy resolution, promising novel insights into both perturbative excitations and non-equilbrium phases in quantum materials.

\begin{acknowledgments}
We gratefully acknowledge A.~Federov, W.~Zhang, S.~Ulonska, and B.~V.~Smith for contributions during commissioning, J.~Wu and S.~Tongay for providing TMD samples and P. Richard for access to software macros. Moreover, we thank R. W. Schoenlein, H.~Ding, and F. Parmigiani for many stimulating discussions. This work was primarily funded by the U.S. Department of Energy (DOE), Office of Science, Office of Basic Energy Sciences, Materials Sciences and Engineering Division under contract no. DE‐AC02‐05CH11231 (Ultrafast Materials Science program KC2203) covering the design and construction of the trARPES setup as well as equilibrium and ultrafast studies (J.H.B, H.W., Y.X., J.M., F.J., L.Z., S.S., A.L., R.A.K.). The cryostat was developed and input into the UHV design provided by the Advanced Light Source (C.J., J.P., Y.D.C., J.D., Z.H.), which is a DOE Office of Science User Facility supported under contract no. DE‐AC02‐05CH11231. J.H.B. gratefully acknowledges a postdoc fellowship from the German Research Foundation (DFG), and J.M. and S.S. acknowledge student fellowships from the German Academic Exchange Service.
\end{acknowledgments}

\vspace{3 mm}

\noindent {\bf Journal Reference:} This is the accepted version of a journal publication and may be downloaded for personal use only. Any other use requires prior permission of the author and AIP Publishing. The article appeared in Rev.~Sci.~Instrum. {\bf 90}, 023105 (2019) and may be found at \href{https://doi.org/10.1063/1.5079677}{https://doi.org/10.1063/1.5079677}.

\bibliography{RSI-XUV-trARPES}
\end{document}